\documentclass[twocolumn,showpacs,preprintnumbers,amsmath,amssymb]{revtex4}

\usepackage[xdvi]{graphicx}
\usepackage{dcolumn}
\usepackage{bm}

\begin{document}

\preprint{EPHOU 02-007}

\title{Analytic Evaluation of the Decay Rate for an Accelerated Proton}

\author{Hisao Suzuki}
\email{hsuzuki@phys.sci.hokdai.ac.jp}
\author{Kunimasa Yamada}
\email{kunimasa@particle.sci.hokudai.ac.jp}

\affiliation{
Department of Physics, Hokkaido University, Sapporo 060-0810, Japan}

\date{\today}

\begin{abstract}
We evaluate the decay rate of the uniformly accelerated proton. We obtain an analytic
expression for inverse beta decay process caused by the acceleration. We evaluate the
decay rate both from the inertial frame and from the accelerated frame where we
should consider thermal radiation by Unruh effect. We explicitly check that the decay
rates obtained in both frame coincide with each other.
\end{abstract}

\pacs{13.30.-a, 04.62.+v, 12.15.Ji, 14.20.Dh}

\maketitle

\section{Introduction}

In 1974 Hawking announced a celebrated result that the quantum thermal radiance
occurs from black hole \cite{hawking}, which is the most interesting result of the
quantum field theory in curved  space-time.
Since then, the study of the field theory has been of much interest. 
It appeared that the quantum radiation is not special phenomenon in strong gravitation but 
analogous effect exists even in the flat space-time \cite{unruh76}.
The effect is caused by the fact that there is no
vacuum which is invariant under the general coordinate transformations although the vacuum
is invariant under the special coordinate transformations. A typical example is the
accelerated observer which feels thermal radiation of the particles depending on
the rate of acceleration.  The effects in the
accelerated system are not hard to be analyzed and is more tractable than the
purely gravitational effects.

This phenomenon in flat space-time was shown by Fulling-Davies-Unruh in
1973, which is now called Unruh effect \cite{fulling73,Davies75}.
A vacuum in rest frame is not a vacuum in 
accelerated systems.
Actually, the vacuum in rest frames is 
a thermal bath in the accelerated frames \cite{sewell82}.
This means that the definition of particle in quantum field theory
depends on the observer and
a strongly curved space-time by the acceleration corresponds to 
a strong gravitation.

Until now, many phenomena which the acceleration gives the interesting
 influence have been shown in the cosmological setting \cite{birrell}.
There is only one experiment about Unruh
effect  by Chen and Tajima \cite{chen99}. 
They found that an electron which is
accelerated by electronic field quivered.
They showed that
the random absorption of quanta from the FDU thermal bath cause 
this quivered.
However, the experiment of decay is not existing.

More recently, the decay of accelerated particles was studied.
In 1997 R. Muller \cite{muller97} estimated the decay rates of the accelerated
particle \cite{caso98}. 
However, it was shown that decay rate is very small in our
experimental setup.

 It was G. E. A Matsas and D. A. T. Vanzella \cite{matsas99,matsas01} who
realized that the decay rate can be used as a "$theoretical \ check$" of Unruh
effect. The most interesting decay is in the case where
we can deal with a particle which doesn't decay in inertial frame but
can in accelerated frame. 
A typical example is for the proton.
In inertial frame the accelerated proton can decay, and we can write
the process as
\begin{equation*}
  \Gamma_{\rm rest}; \;\;\;\;\;p^+\rightarrow n+e^++\nu_e.
\end{equation*}
In the accelerated frame, the observer takes the same acceleration as 
proton then the proton is stable but his space is in a thermal bath of particles.
So the proton should absorb $e^-$ and  $\bar{\nu}$,
and the processes become the inverse $\beta$ decays as
\begin{eqnarray*}
  \Gamma_{\rm acc};\;\;
&& p^++e^-\rightarrow n+\nu,\\
&& p^++\bar{\nu} \;\;\rightarrow n+e^+,\\
&& p^++e^-+\bar{\nu} \rightarrow n.
\end{eqnarray*}
Both of the cross sections should agree with each other because they are 
the probabilities of the identical physical phenomenon 
and it is independ on the frames we use.
However, the calculation is very complicating because
the solutions of the Dirac equation in accelerated frame 
become a special function even if we represent protons by the classical
current and restrict to tree level. 
 Therefore, they restricted the
system to two-dimensions.
They could represent the decay rate analytically for Minkowski frame in the limit
where  neutrino is massless. However, the decay rate for the accelerated frame could not
be evaluated analytically. Therefore,  they showed the agreement numerically for
massless neutrino. This calculation shows that the Unruh effect is inevitable for
the accelerated frame. An interesting point of their approach is that they use a
decay rate, true physical observables, for showing the existence of Unruh effect
although the effect itself was established earlier (see Ref.~\cite{birrell} for
Review).

They used two dimensional setting and treated neutrino for
massless field for simplicity.  However, it is desirable if we can show that decay
rates calculated in both frame are identical even for massive neutrino in four
dimensions because massiveness of neutrino is now proved experimentally.

The main aim of this paper is to show that we can complete their calculation by
showing
 that the decay rate of proton is independent of the frame not by
numerical computation but by analytical computation.  Moreover, We will perform the
evaluation in four dimensional setting and treat neutrino a massive field. It is
interesting that the decay rate can be analyzed analytically.

In Section II we will calculate the cross section of $\beta$ decay
in the inertial frame. We will show that the decay rate can be obtained
analytically in terms of a function which is an analog of Meijer's G-function of two
variables.

In Section III, we will perform the calculation in accelerating frame.
We will be able to check that the resulting function is identical to the one
obtained in inertial frame.

Section IV is devoted to the discussions.

  In Appendix B, we list the explicit form of the function appeared in our
main result.

\section{Inertial frame}

In this section, we analysis the $\beta$ decay of the 
accelerated proton in the inertial frame.
In this frame,
the accelerated proton decay resulting from the acceleration.

In the flat space-time, the solutions of the Dirac equation are simple
but the calculation becomes complicate because of  
the vector current is on the hyperbola. We will mainly follow the notation
appeared in Ref.~\cite{matsas99,matsas01}

\subsection{Accelerated Proton current}

There may be several methods to represent the acceleration. One way is the
followings. We represent the proton as a classical current.
The position of the current should be on the world line of the
accelerated particle, ``hyperbola''.
To do this, we introduce the Rindler coordinates.
Rindler coordinates correspond to the world line of
the uniformly accelerated observers.
The inertial coordinates are shown by $(x^0,x^1,x^2,x^3)$ and
the Rindler coordinates which show acceleration along the 3-axis are
 $(v,x^1,x^2,u)$ with $0<u<\infty$ and $-\infty < v < \infty $.
 Two coordinates are related by
\begin{eqnarray}
 && x^0 = u \sinh{v},\nonumber\\
 && x^1 = x^1,\nonumber\\
 && x^2 = x^2,\nonumber\\
 && x^3 = u \cosh{v}.
\end{eqnarray}
Then the line element is described in the Rindler coordinates by
\begin{equation}
 ds^2 = u^2 dv^2 - (dx^1)^2 - (dx^2)^2 - du^2.
\end{equation}
If some particle uniformly accelerated 
with a proper acceleration $a$, then 
$u=a^{-1}={\rm const.}$ is its world line.
As a result, 
the protons which are accelerated to the axis of $x^3$ 
on $x^1=0, \;x^2=0$ are represented as following classical current 
\begin{equation}
 j^\mu = q u^\mu \delta(x^1) \delta(x^2) \delta(u - a^{-1}),
\end{equation}
where $q$ is a small coupling constant and $u^\mu$ is the 
 four-velocity $u^\mu=(a,0,0,0)$ in Rindler coordinates.
To deal with the proton-decay, nucleons $|n\rangle$ and protons $|p\rangle$
are represented as excited and unexcited states of the nucleon,
respectively.
For Hamiltonian $\hat{H}$, neutron and proton mass $m_n$ and $m_p$,
respectively, we set
\begin{equation}
 \hat{H}|n\rangle = m_n |n\rangle,\;\; \hat{H}|p\rangle = m_p |p\rangle.
\end{equation}
We replace $q$ in last current by the Hermitian monopole
\begin{equation}
 \hat{q}(\tau) \equiv e^{i\hat{H}\tau} \hat{q}(0) e^{-i\hat{H}\tau},
\end{equation}
where $\tau =v/a$ is proper time of the proton.

Then the current four-vector is 
\begin{equation}
 \hat{j}^\mu = \hat{q}(\tau) u^\mu \delta(x^1) \delta(x^2) \delta(u - a^{-1}). 
\end{equation}
This is clear that the particle is on the hyperbola.

\subsection{Fermionic field quantization}

For electrons and
neutrinos, we write the fermionic fields satisfying
the Dirac equation
$(i\gamma^\mu \partial_\mu-m)\psi^{(\pm\omega)}_{\mathbf{k} \sigma}=0$
 as
\begin{equation}
   \hat{\Psi}(x)
    = \sum_{\sigma = \pm} \int_{-\infty}^{\infty}
      d^3 k \left[ \hat{b}_{\mathbf{k} \sigma} 
        \psi^{(+\omega)}_{\mathbf{k}\sigma}(x)
      + \hat{d}_{\mathbf{k} \sigma}^\dag 
      \psi^{(-\omega)}_{\mathbf{k} -\sigma}(x)\right],
\end{equation}
where $x=(x^0,x^1,x^2,x^3)$, $k=(\omega,k^1,k^2,k^3)$,
$\mathbf{x}=(x^1,x^2,x^3)$, $\mathbf{k}=(k^1,k^2,k^3)$.
$\hat{b}_{\mathbf{k} \sigma}$ and $\hat{d}_{\mathbf{k} \sigma}^\dag$ are
annihilation and creation operators of fermions and
antifermions, respectively.
$\mathbf{k}$ and $\sigma$ are momentum and polarization, respectively.
$\psi^{(+\omega)}_{\mathbf{k} \sigma}$ and $\psi^{(-\omega)}_{\mathbf{k}\sigma}$
are positive and negative frequency solutions of the Dirac equation.

By solving the Dirac equation we obtain the orthonormal mode solutions
 \cite{itzykson},
\begin{equation}
 \psi^{(\pm\omega)}_{\mathbf{k} \sigma}(x)
 = \frac{{e^{\mp ik_\mu x^\mu} } }{(2\pi)^\frac{3}{ 2}}
   u^{(\pm\omega)}_\sigma(\bold{k}),
\end{equation}
where
\begin{equation}
 u^{(\pm\omega)}_\sigma(\bold{k})
 = \frac{k_\mu \gamma^\mu \pm m } {\sqrt{\omega (\omega\pm m)}}
   \hat{u}_\sigma
\end{equation}
and
\begin{equation}
 \hat{u}_+
 =
\left[
\begin{array}{c}
1 \\
0 \\
0 \\
0
\end{array}
\right],\;\;\;
 \hat{u}_-
 =
\left[
\begin{array}{c}
0 \\
1 \\
0 \\
0
\end{array}
\right].
\end{equation}

We choose the traditional inner product of the form
\begin{eqnarray}
&& \langle\psi^{(\pm\omega)}_{\mathbf{k} \sigma}(x),
  \psi^{(\pm\omega')}_{\mathbf{k'} \sigma'}(x)\rangle
   \equiv \int_\Sigma d\Sigma_\mu
     \psi^{(\pm\omega)}_{\mathbf{k} \sigma}(x) \gamma^\mu
     \psi^{(\pm\omega')}_{\mathbf{k'} \sigma'}(x)  \nonumber\\ 
  &&\hspace{3.3cm} = \delta^3 (\mathbf{k}-\mathbf{k'}) 
        \delta_{\sigma\sigma'}\delta_{\pm\omega\pm\omega'}
\end{eqnarray}
and the solutions are normalized by this definition,
where $\bar{\psi}\equiv\psi^\dag\gamma^0,\; 
            d\Sigma_\mu\equiv n_\mu d\Sigma$, $n^\mu$ is a unit
vector orthogonal to $\Sigma$ 
and we have chosen $\Sigma$ to be the hypersurface of 
constant $x^0$. 
The creation and annihilation operators satisfy the equal-time 
anti-commutation relations:
\begin{eqnarray}
&&  \left\{ \hat{b}_{\mathbf{k}\sigma},\hat{b}^\dag_{\mathbf{k'}\sigma'} \right\}
 =\left\{ \hat{d}_{\mathbf{k}\sigma},\hat{d}^\dag_{\mathbf{k'}\sigma'} \right\}
 =\delta^3 (\mathbf{k}-\mathbf{k'})\delta_{\sigma\sigma'},\nonumber\\
&&  \left\{ \hat{b}_{\mathbf{k}\sigma},\hat{b}_{\mathbf{k'}\sigma'} \right\}
 =\left\{ \hat{d}_{\mathbf{k}\sigma},
          \hat{d}_{\mathbf{k'}\sigma'} \right\}\nonumber\\
&& \hspace{1cm} =\left\{ \hat{b}_{\mathbf{k}\sigma},
                         \hat{d}_{\mathbf{k'}\sigma'} \right\}
   =\left\{ \hat{b}_{\mathbf{k}\sigma},\hat{d}^\dag_{\mathbf{k'}\sigma'} \right\}
 = 0.
\end{eqnarray}
Using electron $\hat{\Psi}_e$ and neutrino
$\hat{\Psi}_\nu$ fields,
we write the Fermi action by the nucleon current $(2.6)$
\begin{equation}
 \hat{S}_I = \int d^4x\sqrt{-g}\hat{j}_\mu
 (\hat{\bar{\Psi}}_\nu \gamma^\mu \hat{\Psi}_e 
  + \hat{\bar{\Psi}}_e \gamma^\mu \hat{\Psi}_\nu ),
\end{equation}
where the first term is used by inverse $\beta$ decay.

You can see these general formation on any book of Quantum Field Theory
in the inertial frame.
We are ready to start to calculate the cross section of
the accelerated proton in the inertial frame.

\subsection{Calculation of Cross Section}

We are now going to calculate the cross section of the $\beta$ decay
using the current and field in the subsection A. 
The way of deriving the cross section is normal.

First, the vacuum transition amplitude of the proton decay is written by 
\begin{equation}
  {\cal A}^{ p \rightarrow n}
  = \langle n|\otimes\langle e^+_{k_e \sigma_e},\nu_{k_\nu\sigma_\nu}|
  \hat{S}_I |0\rangle\otimes|p\rangle.
\end{equation}
It is straightforward to compute ${\cal A}^{ p \rightarrow n}$
for the Fermi action $\hat{S_I}$  and we obtain 
\begin{eqnarray}
 &&{\cal A}^{ p \rightarrow n}
    = G_F \int_{-\infty}^\infty d(x^0)d(x^3) 
    {e^{i\Delta m\tau}u_\mu\over\sqrt{a^2(x^0)^2+1}}\nonumber\\
 && \times\delta(x^3-\sqrt{(x^0)^2+a^{-2}})
    \langle e^+_{k_e \sigma_e},\nu_{k_\nu\sigma_\nu}|
  \hat{\bar{\Psi}}_\nu \gamma^\mu \hat{\Psi}_e |0\rangle,\;\;\;
\end{eqnarray}
where $\Delta m\equiv m_n-m_p$, $\tau=a^{-1}\sinh^{-1}(ax^0)$ is the
nucleon's proper time, $G_F\equiv \langle p |\hat{q}(0)|n\rangle$
is the Fermi constant.
We can substitute the field and integrate by $ x^3$.
The differential transition rate is
\begin{equation}
 {d^6 {\cal P}^{ p \rightarrow n}_{\rm in}\over d^3k_e d^3k_\nu}
 =\sum_{\sigma_e}\sum_{\sigma_\nu}
  \left|{\cal A}^{ p \rightarrow n}\right|^2.
\end{equation}
We obtain it by integration of proper times $\tau_1$ and $\tau_2$
\begin{widetext}
\begin{eqnarray}
 && {d^6 {\cal P}^{ p \rightarrow n}_{\rm in}\over d^3k_e d^3k_\nu}
   = {G_F^2\over(2\pi)^6} \int_{-\infty}^\infty
   d\tau_1d\tau_2 \:
   u_\mu u_\nu^*
   \sum_{\sigma_e}\sum_{\sigma_\nu}
  \left[\bar{u}^{(+\omega_\nu)}_{\sigma_\nu} 
        \gamma^\mu u^{(-\omega_e)}_{-\sigma_e}
  \right]  
  \left[\bar{u}^{(+\omega_\nu)}_{\sigma_\nu} 
        \gamma^\nu u^{(-\omega_e)}_{-\sigma_e}
  \right]^*
   \nonumber\\
 &&  \hspace{1.7cm}\times 
    \exp i\left[\Delta m(\tau_1-\tau_2) 
    + a^{-1}(\omega_\nu + \omega_e)(\sinh a\tau_1-\sinh a\tau_2)
    - a^{-1}(k^3_\nu + k^3_e)(\cosh a\tau_1 - \cosh a\tau_2)\right].
\end{eqnarray}
\end{widetext}
Now in order to calculate the spin sums,
we use the following standard formula:
\begin{eqnarray}
&& \sum_{\sigma_\alpha}\sum_{\sigma_\beta}
 \left[
       \bar{u}^{(\pm\omega)}_{\sigma_\alpha}\Gamma_1 
             u^{(\pm\omega)}_{\sigma_\beta}
 \right]
 \left[
       \bar{u}^{(\pm\omega)}_{\sigma_\alpha}\Gamma_2 
             u^{(\pm\omega)}_{\sigma_\beta}
 \right]^*\nonumber\\
&& = {\rm Tr}
  \left[
   \Gamma_1
     \sum_{\sigma_\alpha}u^{(\pm\omega)}_{\sigma_\alpha}
                    \bar{u}^{(\pm\omega)}_{\sigma_\alpha} 
    \gamma^0\Gamma^\dag_2\gamma^0
     \sum_{\sigma_\beta}u^{(\pm\omega)}_{\sigma_\beta}
                   \bar{u}^{(\pm\omega)}_{\sigma_\beta}
  \right],\nonumber\\
\end{eqnarray}
where $\alpha$ and $\beta$ represent $e$ or $\nu$.

The completeness relations can be written as
\begin{equation}
 \sum_{\sigma_\alpha}
       u^{(\pm\omega_\alpha)}_{\sigma_\alpha}(\mathbf{k}_\alpha)
 \bar{u}^{(\pm\omega_\alpha)}_{\sigma_\alpha}(\mathbf{k}_\alpha)
      ={1\over 2\omega_\alpha}({k_\alpha}_\mu \gamma^\mu \pm m_\alpha)
\end{equation}
and we introduce $s$ and $\xi$ by
\begin{equation}
 \tau_1=s+{\xi\over2},\;\;\tau_2=s-{\xi\over2}.
\end{equation}
By using the spin sum
\begin{eqnarray}
 &&  \sum_{\sigma_e}\sum_{\sigma_\nu}
  \left[
   \bar{u}^{(+\omega_\nu)}_{\sigma_\nu} 
   \gamma^\mu u^{(-\omega_e)}_{-\sigma_e}
  \right]  
  \left[
   \bar{u}^{(+\omega_\nu)}_{\sigma_\nu} 
   \gamma^\nu u^{(-\omega_e)}_{-\sigma_e}
  \right]^*  \nonumber\\
 &&  = {1\over\omega_e\omega_\nu}
  \bigl[
   (\omega_e\omega_\nu+k^3_ek^3_\nu)\cosh2as\nonumber\\
 && \;\;\;\;  -(\omega_ek^3_\nu+k^3_e\omega_\nu)\sinh2as
    -m_em_\nu\cosh a\xi
    \bigr] \nonumber\\
 && \;\;\;\;\;\;\;+(\mbox{odd function of $k^1_\alpha$ and
  $k^2_\alpha$})
  \nonumber\\[5mm]
\end{eqnarray}
and the change of variables
\begin{equation}
 {k'}^1_\alpha=k^1_\alpha,\;
 {k'}^2_\alpha=k^2_\alpha,\;
 {k'}^3_\alpha=-\omega_\alpha \sinh as+k^3_\alpha \cosh as,
\end{equation}
we can obtain the differential transition rate 
\begin{eqnarray}
 && {1\over T}{d^6 {\cal P}^{ p \rightarrow n}_{\rm in}\over d^3k'_e d^3k'_\nu}
   ={G_F^2\over(2\pi)^6} {1\over\omega'_e\omega'_\nu}\int_{-\infty}^\infty
   d\xi 
   e^{i
    [\Delta m \xi+{\omega'_e+\omega'_\nu\over 2a}
    \sinh{a\xi\over2}]}
    \nonumber\\ 
 &&  \hspace{2.3cm}\times(\omega'_e\omega'_\nu+{k'}^3_e{k'}^3_\nu-m_em_\nu
     \cosh a\xi),\nonumber\\[5mm]
\end{eqnarray}
where $T\equiv \int^\infty_{-\infty}ds$.
To integrate by $\xi$ we redefine new variable by
\begin{equation}
 \lambda\equiv e^{a\xi\over 2}.
\end{equation}
And we use the following notations 
\begin{equation}
 \widetilde{k}_\alpha\equiv{k'_\alpha\over a},\;\;
 \widetilde{\omega}_\alpha\equiv{\omega_\alpha\over a},\;\;
 \widetilde{m}_\alpha\equiv{m_\alpha\over a},\;\;
 \widetilde{\Delta m}\equiv{\Delta m\over a}.
\end{equation} 

Then the cross section is given in the form: 
\begin{eqnarray}
 && \Gamma^{p\rightarrow n}_{\rm in} 
   = {1\over T}\int 
     d {\cal P}^{ p \rightarrow n}_{\rm in}  
     \nonumber\\ 
 &&\hspace{9mm}={a^5G_F^2\over 2^5\pi^6} \int_{-\infty}^\infty
   {d^3\widetilde{k}_e d^3\widetilde{k}_\nu
    \over
    \tilde{\omega}_e\tilde{\omega}_\nu}
   d\lambda \:
    e^{i(\widetilde{\omega}_e+\widetilde{\omega}_\nu)
               (\lambda-{1\over\lambda})}
   \nonumber\\ 
 && \hspace{1.3cm}\times\lambda^{2i\widetilde{\Delta m}-1}
   \left[\widetilde{\omega}_e \widetilde{\omega}_\nu-{1\over2}
         \widetilde{m}_e \widetilde{m}_\nu (\lambda^2+{1\over\lambda^2})
   \right].\;\;\;\nonumber\\[5mm]
\end{eqnarray}
The integral of $\lambda$ can be readily expressed as modified
Bessel  to find 
\begin{eqnarray}
 && \Gamma^{p\rightarrow n}_{\rm in} 
   ={2^2a^5G_F^2\over\pi^6e^{\pi\widetilde{\Delta m}}}
 \int_0^\infty
   d^3\widetilde{k}_e d^3\widetilde{k}_\nu
 \biggl[K_{2i\widetilde{\Delta m}}
  \left({{}\over{}}2(\widetilde{\omega}_e+\widetilde{\omega}_\nu)\right)
 \nonumber\\
 && \hspace{1.5cm} +{1\over2}{\widetilde{m}_e\widetilde{m}_\nu\over
     \widetilde{\omega}_e\widetilde{\omega}_\nu}
     \left\{
      K_{2i\widetilde{\Delta m}+2}
      \left(
       {{}\over{}}2(\widetilde{\omega}_e+
                     \widetilde{\omega}_\nu)
      \right) 
     \right.
   \nonumber\\
 && \hspace{3.2cm}
     \left.
      + K_{2i\widetilde{\Delta m}-2}
      \left(
       {{}\over{}}2(\widetilde{\omega}_e+
                     \widetilde{\omega}_\nu)
      \right)
     \right\}
    \biggr].\nonumber\\[5mm]
    \label{eq:1}
\end{eqnarray}
However, this form is difficult to integrate by $\widetilde{k}_e$ and
$\widetilde{k}_\nu$.
Therefore, we use the integration formula of modified Bessel
\begin{equation}
 K_\mu(z)={1\over2}\int_{C_1} {ds\over 2\pi i}\Gamma(-s)\Gamma(-s-\mu)
       \left({z\over2}\right)^{2s+\mu}.
 \label{k}
\end{equation}
It is not hard to show this formula in complex $s$ plane
by picking the residues of $\Gamma(-s)$ and $\Gamma(-s-\mu)$(see
Appendix A).
To evaluate Eq.~({\ref{eq:1}}) we use this formals and
after some simple shift of variable we have
\begin{widetext} 
\begin{eqnarray} 
  && \Gamma^{p\rightarrow n}_{\rm in} 
={2a^5G_F^2\over\pi^6e^{\pi\widetilde{\Delta m}}}
 \int_0^\infty
   d^3\widetilde{k}_e d^3\widetilde{k}_\nu
   \int_{C_1} {ds\over2\pi i}
   \left(\sqrt{\tilde{k}_e^2+\tilde{m}_e^2}
        +\sqrt{\tilde{k}_\nu^2+\tilde{m}_\nu^2}\right)
        ^{2s+2i\widetilde{\Delta m}}
 \nonumber\\
&& \hspace{1.3cm}\times
   \Biggl[
     \Gamma(-s)\Gamma(-s-2i\widetilde{\Delta m})
    +{1\over2}{\widetilde{m}_e\widetilde{m}_\nu\over
            \widetilde{\omega}_e\widetilde{\omega}_\nu}
  \left\{
     \Gamma(-s+1)\Gamma(-s-2i\widetilde{\Delta m}-1)
    +\Gamma(-s-1)\Gamma(-s-2i\widetilde{\Delta m}+1) 
  \right\} \Biggr],\;\;\;\;\;
\label{eq:one}
\end{eqnarray}
where the contour $C_1$ must be chose so that all poles of
 $\Gamma(-s)\Gamma(-s-\mu)$ are picked up and
 must be selected as $k$ integration don't infinity.
But, this form of integration is still not simple for $\widetilde{k}_e$ and
$\widetilde{k}_\nu$.

In order to perform the integration, we use the following expansion formula 
\begin{equation}
 (A+B)^z = \int_{C_2} {dt\over2\pi i}{\Gamma(-t)\Gamma(t-z)\over \Gamma(-z)}
           A^{t+z}B^t\label{k2}
\end{equation}
to integrate easily by $\widetilde{k}_e$ and
$\widetilde{k}_\nu$, where the contour $C_2$
is the path separating the poles of $\Gamma(-t)$ 
from those of $\Gamma(t-z)$ (see Appendix A).

Now we can rewrite Eq.~(\ref{eq:one}) as
\begin{eqnarray}
 && \Gamma^{p\rightarrow n}_{\rm in}
={a^5G_F^2 \over 2^5\pi^{7\over2}e^{\pi\widetilde{\Delta m}}}
    \int_{C_s}{ds\over2\pi i}\int_{C_t}{dt\over2\pi i}
    {
     (\tilde{m}_e^2)^s
     (\tilde{m}_\nu^2)^t
     \over
     \Gamma(-s-t+3)
     \Gamma(-s-t+{7\over2})
    }\nonumber\\
 && \hspace{1.3cm}\times
    \Biggl[
     \left|\Gamma(-s-t+i\widetilde{\Delta m}+3)\right|^2
     \Gamma(-s)
     \Gamma(-t)
     \Gamma(-s+2)
     \Gamma(-t+2)   
     \nonumber\\
 && \hspace{1.8cm}
    +{\rm Re}\left\{
          \Gamma(-s-t+i\widetilde{\Delta m}+2)
          \Gamma(-s-t-i\widetilde{\Delta m}+4)
        \right\}
     \Gamma(-s+{1\over2})
     \Gamma(-t+{1\over2})
     \Gamma(-s+{3\over2})
     \Gamma(-t+{3\over2})   
     \Biggr]
\label{result1}
\end{eqnarray}
\end{widetext} 
where the contour $C_s$ and $C_t$ is 
the path which picks up all poles of Gamma functions in $s$ and $t$
complex planes, respectively.

This is the two-dimensional analog of Meijer's G-function \cite{HTF}. The
explicit form can be obtained by evaluating contour integral. We list the
results in the Appendix B.

\section{Accelerated Frame}

In this section, we are going to analyze the same physical phenomenon view
of accelerated system. 

\subsection{Fermionic field quantization}

The Dirac equation in curved space-time is written by
\begin{equation}
 (i\gamma^\mu_R \widetilde{\nabla}_\mu-m)
 \psi^{(\omega)}_{\mathbf{w}\sigma}(x)=0,
\end{equation}
where
\begin{eqnarray}
  && x=(v,x^1,x^2,u),\nonumber\\
  && \mathbf{w}=(\omega,k^1,k^2),\;\;
     \mathbf{x}=(x^1,x^2,u),\nonumber\\
 &&  (e_0)^\mu=u^{-1}\delta^\mu_0,\;(e_i)^\mu=\delta^\mu_i,\nonumber\\
  && \gamma^\mu_R \equiv (e_\nu)^\mu\gamma^\nu,\nonumber\\
 && \widetilde{\nabla}_\mu\equiv\partial_\mu+
     {1\over8}[\gamma^\alpha,\gamma^\beta]
 (e_\alpha)^\lambda\nabla_\mu(e_\beta)_\lambda.
\end{eqnarray}
In the Rindler coordinates, the equation becomes
\begin{equation}
 i{\partial\psi^{(\omega)}_{\mathbf{w}\sigma}(x)\over\partial v}
 = \left(\gamma^0mu-{i\alpha^3\over2}-iu\alpha^i\partial_i\right)
  \psi^{(\omega)}_{\mathbf{w}\sigma}(x),
\end{equation}
where $\alpha^i\equiv \gamma^0\gamma^i $.

The fermionic field can be expanded as
\begin{eqnarray}
 &&  \hat{\Psi}(x) =
    \sum_{\sigma=\pm} \int_0^{\infty}d\omega 
      \int_{-\infty}^\infty d^2k
      \left[ \hat{b}_{\mathbf{w}\sigma} 
          \psi^{(+\omega)}_{\mathbf{w} \sigma}
      + \hat{d}_{\mathbf{w} \sigma}^\dag 
      \psi^{(-\omega)}_{\mathbf{w} -\sigma}\right].\nonumber\\
\end{eqnarray}
We will the solution in the form
\begin{equation}
 \psi^{(\omega)}_{\mathbf{w}\sigma}
 =\bar{f}_{\mathbf{w}\sigma}(\mathbf{x})e^{-i\omega v/a}
\end{equation}
for $-\infty < \omega< \infty$.

The function $\bar{f}_{\mathbf{w}\sigma}$ is eigenstate of Hamiltonian as
\begin{equation}
 \hat{H}\bar{f}_{\mathbf{w}\sigma} = \omega \bar{f}_{\mathbf{w}\sigma},
\end{equation}
where
\begin{equation}
 \hat{H}= a\left[mu\gamma^0-{i\alpha^3\over2}-iu\alpha^i\partial_i\right].
\end{equation}
We denote two-component spinors $\chi_a$ by
\begin{equation}
 \bar{f}_{\mathbf{w}\sigma}(\mathbf{x})
 \equiv
\left[
\begin{array}{c}
 \bar{\chi}_1(\mathbf{x},\mathbf{w}) \\
 \bar{\chi}_2(\mathbf{x},\mathbf{w})
\end{array}
\right].
\end{equation}
then we find that the Dirac equation takes the form
\begin{subequations}
\label{eq:whole}
\begin{eqnarray}
 && \delta^{ij}u\partial_i(u\partial_j\chi_1)
 = \left[m^2u^2+{1\over4}-{\tilde{\omega}^2}\right]\chi_1
  - i\tilde{\omega}\sigma^3\chi_2,\\
 && \delta^{ij}u\partial_i(u\partial_j\chi_2)
 = \left[m^2u^2+{1\over4}-{\tilde{\omega}^2}\right]\chi_2
  - i\tilde{\omega}\sigma^3\chi_1,\;\;\;\;\;\;\;
\end{eqnarray}
\end{subequations}
where $i$ and $j$ run over 1, 2, 3.
By now, for squared equations we used $f_{\mathbf{w}\sigma}$ , $\chi_1$
 and $\chi_2$.

To simplify these equations, we introduce the functions $\phi^\pm\equiv
\chi_1\mp\chi_2$ and
we can define $\xi^\pm$ and $\zeta^\pm$ through
\begin{equation}
 \phi^\pm(\mathbf{x},\mathbf{w})
 \equiv
\left[
\begin{array}{c}
\xi^\pm(\mathbf{x},\mathbf{w}) \\
\zeta^\pm(\mathbf{x},\mathbf{w})
\end{array}
\right].
\end{equation}
Given this definition, we can separate the equation for  $\xi$ and
$\zeta$ in the form:
\begin{subequations}
\begin{eqnarray}
 && \delta^{ij}u\partial_i(u\partial_j\xi^\pm)
 = \left[m^2u^2+\left(i\tilde{\omega}\pm{1\over2}\right)^2\right]\xi^\pm,\\
 && \delta^{ij}u\partial_i(u\partial_j\zeta^\pm)
 = \left[m^2u^2+\left(i\tilde{\omega}\mp{1\over2}\right)^2\right]\zeta^\pm.
\end{eqnarray}
\end{subequations}
Here, we introduce $\ell$ as  $\ell^2 = (k^1)^2 + (k^2)^2 + m^2$ by which the
equation can be written as
\begin{eqnarray}
   [\partial_1^2+\partial_2^2-m^2]\xi^\pm
 &=& {1\over u^2}\biggl[-u\partial_3(u\partial_3)
   +\left(i\tilde{\omega}\pm{1\over2}\right)^2\biggr]\xi^\pm\nonumber\\
 &\equiv& -\ell^2 \xi^\pm.
\end{eqnarray}
The solution can be written in the form 
\begin{equation}
 \xi^\pm = A^\pm(\mathbf{w}){\cal X}(x^1,x^2,k^1,k^2)
  {\cal M}_{\mathbf{w}}^\pm(u).  
\end{equation}
This leads to the following decoupled equations
\begin{eqnarray}
 && [\partial_1^2+\partial_2^2-m^2]{\cal X}(x^1,x^2,k^1,k^2)=0,\\
 && \left[{\partial^2\over\partial(\ell u)^2}
       +{1\over \ell u}{\partial\over\partial(\ell u)}
       -1-\left({i\tilde{\omega}\mp{1\over2}
       \over \ell u}\right)^2\right]{\cal M}_{\mathbf{w}}^\pm(u) = 0.
    \nonumber\\
\end{eqnarray}
We can readily solve the equation in the form
\begin{eqnarray}
 && {\cal X}(x^1,x^2,k^1,k^2)=e^{ik_ax^a},\\
 && {\cal M}_{\mathbf{w}}^\pm(u) = K_{i\tilde{\omega}\mp{1\over2}}(\ell u),
\end{eqnarray}
where the index $a$ run over 1 and 2.

Using the arbitrary function $A^\pm$ and $B^\pm$ of $k$, we find the most general
solution can be written as 
\begin{subequations}
\begin{eqnarray}
 && \xi^\pm = A^\pm(\mathbf{w})e^{ik_ax^a}
    K_{i\tilde{\omega}\mp{1\over2}}(\ell u),\\
 && \zeta^\pm = B^\pm(\mathbf{w})e^{ik_ax^a}
    K_{i\tilde{\omega}\pm{1\over2}}(\ell u),
\end{eqnarray}
\end{subequations}
which implies
\begin{eqnarray}
 && f_{\mathbf{w}\sigma}
  ={1\over2}
\left[
\begin{array}{c}
  \;\;\;\xi^++\xi^-\\
  \;\;\;\zeta^++\zeta^- \\
 -\xi^++\xi^- \\
 -\zeta^++\zeta^-
\end{array}
\right]\nonumber\\
&&\hspace{6mm} ={e^{ik_ax^a}\over2}
\left[
\begin{array}{c}
  \;\;\;
  A^+K_{i\tilde{\omega}-{1\over2}}(\ell u)
 +A^-K_{i\tilde{\omega}+{1\over2}}(\ell u)\\
  \;\;\;
  B^+K_{i\tilde{\omega}+{1\over2}}(\ell u)
 +B^-K_{i\tilde{\omega}-{1\over2}}(\ell u)\\
 -A^+K_{i\tilde{\omega}-{1\over2}}(\ell u)
 +A^-K_{i\tilde{\omega}+{1\over2}}(\ell u)\\
 -B^+K_{i\tilde{\omega}+{1\over2}}(\ell u)
 +B^-K_{i\tilde{\omega}-{1\over2}}(\ell u)
\end{array}
\right].\;\;\;\;\;\;
\end{eqnarray}
It is easy to see that $f_{\mathbf{w}\sigma}$ satisfies
\begin{eqnarray}
 && \left[i\alpha^i\partial_i-m\gamma^0+\left({i\alpha^3\over2}
  +\tilde{\omega}\right){1\over u}\right]\nonumber\\
 && \hspace{5mm}\times\left[i\alpha^j\partial_j-m\gamma^0+\left({i\alpha^3\over2}
  -\tilde{\omega}\right){1\over u}\right]f_{\mathbf{w}\sigma}=0.
\end{eqnarray}
Therefore, the solutions of Dirac equation are simply given by
\begin{equation}
 \bar{f}_{\mathbf{w}\sigma}\equiv
 \left[i\alpha^i\partial_i-m\gamma^0+\left({i\alpha^3\over2}
  -\tilde{\omega}\right){1\over u}\right]f_{\mathbf{w}\sigma}.
\end{equation}
If we set 
\begin{equation}
 A^+=B^-=0
\end{equation}
we can find a normalized $A^-$ and $B^+$ by letting
\begin{equation}
 A^-=B^+=
 \left[\cosh\tilde{\omega}\pi
 \over \pi\tilde{\ell }\right]
  ^{1\over2}\equiv N,
\end{equation}
where
\begin{equation}
 \widetilde{\ell}\equiv{\ell\over a},
\end{equation} 
which is normalized with respect to the inner product 
\begin{eqnarray}
 && \langle\psi^{(\omega)}_{\mathbf{w}\sigma}(x),
  \psi^{(\omega')}_{\mathbf{w'}\sigma'}(x)\rangle
     \equiv  \int_\Sigma d\Sigma_\mu
     \psi^{(\omega)}_{\mathbf{w}\sigma}(x) \gamma^\mu_R
      \psi^{(\omega')}_{\mathbf{w'}\sigma'}(x) \nonumber\\ 
  && \hspace{3cm}=  \delta^3(\mathbf{w}-\mathbf{w'}) \delta_{\sigma\sigma'},
\end{eqnarray}
where $\bar{\psi}\equiv\psi^\dag\gamma^0$
and $\Sigma$ is set to be $v={\rm const.}$.
In this way, we find concrete form of fermionic field is written by
\begin{equation}
 \psi^{(\omega)}_{\mathbf{w} \sigma}(x)
 = {e^{-i\omega v/a+ik_ax^a} \over (2\pi)^{3\over2}}
   u_\sigma^{(\omega)}(u,\mathbf{w}),
\end{equation}
where
\begin{eqnarray}
 && u_\sigma^{(\omega)}(u,\mathbf{w})\nonumber\\ 
 &&=  N\gamma^0\left[(\tilde{k}_a\gamma^a + \tilde{m})
   K_{i\tilde{\omega}+{1\over2}}(\ell u)
   +i\tilde{\ell }\gamma^3K_{i\tilde{\omega}-{1\over2}}(\ell u)
\right]\hat{u}_\sigma\nonumber\\ 
\end{eqnarray}
and
\begin{equation}
 \hat{u}_+
 =
\left[
\begin{array}{c}
1 \\
0 \\
1 \\
0
\end{array}
\right],\;\;\;
\hat{u}_- 
 =
\left[
\begin{array}{c}
\;\;\;0 \\
\;\;\;1 \\
\;\;\;0 \\
-1
\end{array}
\right].
\end{equation}
The creation and annihilation operators should obey
\begin{eqnarray}
&&  \left\{ \hat{b}_{\mathbf{w}\sigma},\hat{b}^\dag_{\mathbf{w'}\sigma'} \right\}
 =\left\{ \hat{d}_{\mathbf{w}\sigma},\hat{d}^\dag_{\mathbf{w'}\sigma'}
 \right\}
 =\delta^3(\mathbf{w}-\mathbf{w'})\delta_{\sigma\sigma'},\nonumber\\
 && \left\{ \hat{b}_{\mathbf{w}\sigma},\hat{b}_{\mathbf{w'}\sigma'} \right\}
 =\left\{ \hat{d}_{\mathbf{w}\sigma},\hat{d}_{\mathbf{w'}\sigma'} \right\}
 \nonumber\\
 &&\hspace{1.5cm}=\left\{ \hat{b}_{\mathbf{w}\sigma},\hat{d}_{\mathbf{w'}\sigma'} \right\}
 =\left\{ \hat{b}_{\mathbf{w}\sigma},\hat{d}^\dag_{\mathbf{w'}\sigma'} \right\}
 = 0.
\end{eqnarray}

\subsection{Calculation of Cross Section}

The process of $\beta$ decay in the accelerated frame looks
very different from that in rest frame.
In this case, a proton is stable 
but the whole space is FDU thermal bath characterized by a 
temperature $T=a/2\pi$.
Therefore, proton absorbs $e^-$ and $\bar{\nu}$ from FDU thermal bath and
do emit $e^+$ and $\nu$.
Three processes are possible through the precesses: 
\begin{eqnarray*}
&& (\rm i)\;\;\;p^+ + e^-\rightarrow n+\nu,\\
&& (\rm ii)\;\;p^+ + \bar{\nu}\;\;\rightarrow n + e^+,\\
&& (\rm iii)\;p^+ + e^- + \bar{\nu}\rightarrow n.
\end{eqnarray*}
The transition rate is a combination of them. 

Formally, we can calculate the cross sections 
by same way of the inertial system but we have to deal with three
processes.

The transition amplitudes are written by
\begin{subequations}
\begin{eqnarray}
  && {\cal A}^{ p \rightarrow n}_{(\rm i)}
    = \langle n|\otimes\langle\nu_{k_\nu\sigma_\nu}|
      \hat{S}_I |e^-_{k_{e^-}\sigma_{e^-}}\rangle\otimes|p\rangle\nonumber\\
  &&  \hspace{1cm}={G_F\over a} \int_{-\infty}^\infty dv e^{i\Delta mv}
  \langle\nu_{k_\nu\sigma_\nu}|
  \hat{\bar{\Psi}}_\nu \gamma^0 \hat{\Psi}_e
  |e^-_{k_{e^-} \sigma_{e^-}}\rangle\nonumber\\
  && \hspace{1cm}={G_F\over 2\pi}\delta(\omega_{e^-}-\omega_{\nu}-\Delta m)
       \bar{u}^{(\omega_\nu)}_{\sigma_\nu}
       \gamma^0 u^{(\omega_{e^-})}_{\sigma_{e^-}},\\[8mm]
  && {\cal A}^{ p \rightarrow n}_{(\rm ii)}
    = {G_F\over 2\pi}\delta(\omega_{\bar{\nu}}-\omega_{e^+}-\Delta m)
       \bar{u}^{(\omega_{e^+})}_{\sigma_{e^+}}
       \gamma^0 u^{(\omega_{\bar{\nu}})}_{\sigma_{\bar{\nu}}},\\[8mm]
  && {\cal A}^{ p \rightarrow n}_{(\rm iii)}
    = {G_F\over 2\pi}
       \delta(\omega_{e^-}+\omega_{\bar{\nu}}-\Delta m)
       \bar{u}^{(-\omega_{\bar{\nu}})}_{-\sigma_{\bar{\nu}}}
       \gamma^0 
       u^{(\omega_{e^-})}_{\sigma_{e^-}},\\[8mm]\nonumber
\end{eqnarray}
\end{subequations}
where $\hat{S}_I$ is replaced $\gamma^\mu$ in inertial frame with
$\gamma^\mu_R$ and $u^\mu=(a,0,0,0)$.

We assume that the observer is in the thermal bath. 
 We attach the fermionic thermal factor for each process.
Then the differential transition rate per absorbed and emitted particle
energies for each processes are written as
\begin{subequations}
\begin{eqnarray}
 &&  {1\over T}{d^6 {\cal P}^{ p \rightarrow n}_{(\rm i)}
     \over d\omega_{e^-} d\omega_\nu  d^2k_{e^-} d^2k_\nu} \nonumber\\
 &&= {1\over T}\sum_{\sigma_{e^-}}\sum_{\sigma_\nu}
     \left|{\cal A}^{ p \rightarrow n}_{(\rm i)}\right|^2  
     n_F(\omega_{e^-})[1-n_F(\omega_{\nu})]
     \nonumber\\
 &&= {G_F^2\over 2^5\pi^3}
     {\left|\bar{u}^{(\omega_\nu)}_{\sigma_\nu}
       \gamma^0 u^{(\omega_{e^-})}_{\sigma_{e^-}}\right|^2
       \delta(\omega_{e^-}-\omega_\nu-\Delta m)
      \over e^{\pi\widetilde{\Delta m}}\cosh\tilde{\omega}_{e^-}\pi 
             \cosh\tilde{\omega}_\nu\pi}
      ,\;\;
\end{eqnarray}

\begin{eqnarray}
 && {1\over T}{d^6 {\cal P}^{ p \rightarrow n}_{(\rm ii)}
     \over d\omega_{\bar{\nu}} d\omega_{e^+} d^2k_{\bar{\nu}} d^2k_{e^+}} \nonumber\\
 &&= {1\over T}\sum_{\sigma_{e^+}}\sum_{\sigma_{\bar{\nu}}}
     \left|{\cal A}^{ p \rightarrow n}_{(\rm ii)}\right|^2  
     n_F(\omega_{\bar{\nu}})[1-n_F(\omega_{e^+})]
     \nonumber\\
 &&= {G_F^2\over 2^5\pi^3}
     {\left|
       \bar{u}^{(\omega_{e^+})}_{\sigma_{e^+}}
       \gamma^0 u^{(\omega_{\bar{\nu}})}_{\sigma_{\bar{\nu}}}
     \right|^2
      \delta(\omega_{\bar{\nu}}-\omega_{e^+}-\Delta m)
      \over e^{\pi\widetilde{\Delta m}}\cosh\tilde{\omega}_{\bar{\nu}}\pi 
             \cosh\tilde{\omega}_{e^+}\pi}
      ,\;\;
\end{eqnarray}
\begin{eqnarray}
 && {1\over T}{d^6 {\cal P}^{ p \rightarrow n}_{(\rm iii)}
    \over d\omega_{e^-} d\omega_{\bar{\nu}}d^2k_{e^-} d^2k_{\bar{\nu}}} \nonumber\\
 &&= {1\over T}\sum_{\sigma_{e^-}}\sum_{\sigma_{\bar{\nu}}}
     \left|{\cal A}^{ p \rightarrow n}_{(\rm iii)}\right|^2  
     n_F(\omega_{e^-})n_F(\omega_{\bar{\nu}})
     \nonumber\\
 &&= {G_F^2\over 2^5\pi^3}
     {\left|
       \bar{u}^{(-\omega_{\bar{\nu}})}_{-\sigma_{\bar{\nu}}}
       \gamma^0 
             u^{(\omega_{e^-})}_{\sigma_{e^-}}
      \right|^2
      \delta(\omega_{e^-}+\omega_{\bar{\nu}}-\Delta m)
      \over e^{\pi\widetilde{\Delta m}}\cosh\tilde{\omega}_{e^-}\pi 
             \cosh\tilde{\omega}_{\bar{\nu}}\pi},\;\;
\end{eqnarray}
\end{subequations}
where
\begin{equation}
 n_F(\omega)\equiv{1\over 1+e^{2\pi\tilde{\omega}}}
\end{equation}
is the fermionic thermal factor 
and $T=2\pi\delta(0)$ is total proper time of the proton.

We can obtain the completeness relations by
\begin{widetext}
\begin{eqnarray}
 && \sum_\sigma u^{(\omega)}_\sigma(u,\mathbf{w})
             \bar{u}^{(\omega)}_\sigma(u,\mathbf{w})
 =  N^2\gamma^0\left[
     2\tilde{\ell }^2
   \left| K_{i\tilde{\omega}+{1\over2}}(\tilde{\ell })\right|^2
  +i\tilde{m}\tilde{\ell }\left\{
                  (\gamma^3-\gamma^0)
                   K^2_{i\tilde{\omega}+{1\over2}}(\tilde{\ell })                 
                 +(\gamma^3+\gamma^0)
                   K^2_{i\tilde{\omega}-{1\over2}}(\tilde{\ell })  
              \right\}
         \right].\;\;
\end{eqnarray}
Using these relations, a direct calculation yields 
the spin sum for the process ({\rm i}) as
\begin{eqnarray}
 && \sum_{\sigma_{e^-}}\sum_{\sigma_\nu}
  \left|\bar{u}^{(\omega_\nu)}_{\sigma_\nu} 
        \gamma^0 
              u^{(\omega_{e^-})}_{\sigma_{e^-}}\right|^2 \nonumber\\
 &&= {2^2\over \pi^4} \cosh\tilde{\omega}_{e^-}\pi
                        \cosh\tilde{\omega}_\nu\pi
      \left[\tilde{\ell }_{e^-} \tilde{\ell }_\nu 
\left| K_{i\tilde{\omega}_{e^-}+{1\over2}}(\tilde{\ell }_{e^-}) 
       K_{i\tilde{\omega}_\nu+{1\over2}}(\tilde{\ell }_\nu)\right|^2\right.    
     +\tilde{m}_e \tilde{m}_\nu
     \left.{\rm Re}
      \left\{
       K^2_{i\tilde{\omega}_{e^-}+{1\over2}}(\tilde{\ell }_{e^-})
       K^2_{i\tilde{\omega}_\nu-{1\over2}}(\tilde{\ell }_\nu)
      \right\}\right].
\label{eq:2}
\end{eqnarray}
We can perform the analogous calculation for the processes $({\rm ii})$ and $({\rm
iii})$ .

According to  Eq.~({\ref{eq:2}}),
the differential transition rate of the process $({\rm i})$ is given by
\begin{eqnarray}
 && {1\over T}{d^6 {\cal P}^{ p \rightarrow n}_{(\rm i)}
    \over d\omega_{e^-} d\omega_\nu  d^2k_{e^-} d^2k_\nu}\nonumber\\ 
 &&= {G_F^2\over 2^3\pi^7
     e^{\pi\widetilde{\Delta m}}}
     \delta(\tilde{\omega}_{e^-}-\tilde{\omega}_\nu
            -\widetilde{\Delta m})   
      \left[\tilde{\ell }_{e^-} \tilde{\ell }_\nu 
      \left| K_{i\tilde{\omega}_{e^-}+{1\over2}}(\tilde{\ell }_{e^-}) 
             K_{i\tilde{\omega}_\nu+{1\over2}}(\tilde{\ell }_\nu)\right|^2  
     +\tilde{m}_e \tilde{m}_\nu
     {\rm Re}
      \left\{
       K^2_{i\tilde{\omega}_{e^-}+{1\over2}}(\tilde{\ell }_{e^-})
       K^2_{i\tilde{\omega}_\nu-{1\over2}}(\tilde{\ell }_\nu)
      \right\}\right]. \;\;\;\;\;
\end{eqnarray}
By using them, the cross sections for each processes can be obtained as 
\begin{subequations}
\begin{eqnarray}
 && \Gamma^{p\rightarrow n}_{(\rm i)}
   = {1\over T}\int
     d {\cal P}^{ p \rightarrow n}_{(\rm i)} 
     \nonumber\\ 
 &&\hspace{9mm}= {2G_F^2\over \pi^7e^{\pi\widetilde{\Delta m}}}
       \int_{\widetilde{\Delta m}}^\infty 
       d\tilde{\omega}_{e^-}
       \Biggl[
       \int_0^\infty
       d^2\widetilde{k}_{e^-} 
       \tilde{\ell }_{e^-} 
 \left|K_{i\tilde{\omega}_{e^-}+{1\over2}}(\tilde{\ell }_{e^-})\right|^2
       \int_0^\infty
       d^2\widetilde{k}_\nu  
       \tilde{\ell }_\nu 
 \left|K_{i(\tilde{\omega}_{e^-}-\widetilde{\Delta m})
        +{1\over2}}(\tilde{\ell }_\nu)\right|^2  
       \nonumber\\
 && \hspace{5cm}+\tilde{m}_e \tilde{m}_\nu {\rm Re}
      \left\{
       \int_0^\infty
       d^2\widetilde{k}_{e^-} 
       K^2_{i\tilde{\omega}_{e^-}+{1\over2}}(\tilde{\ell }_{e^-})
       \int_0^\infty
       d^2\widetilde{k}_\nu  
       K^2_{i(\tilde{\omega}_{e^-}-\widetilde{\Delta m})
        -{1\over2}}(\tilde{\ell }_\nu)
      \right\}\Biggr],
\end{eqnarray}
\begin{eqnarray}
 && \Gamma^{p\rightarrow n}_{(\rm ii)} 
   = {2G_F^2\over \pi^7e^{\pi\widetilde{\Delta m}}}
       \int_0^\infty 
       d\tilde{\omega}_{e^+}
       \Biggl[
       \int_0^\infty
       d^2\widetilde{k}_{e^+} 
       \tilde{\ell }_{e^+} 
      \left|K_{i\tilde{\omega}_{e^+}+{1\over2}}(\tilde{\ell }_{e^+})\right|^2
       \int_0^\infty
       d^2\widetilde{k}_{\bar{\nu}}  
       \tilde{\ell }_{\bar{\nu}}
      \left| K_{i(\tilde{\omega}_{e^+}-\widetilde{\Delta m})
        +{1\over2}}(\tilde{\ell }_{\bar{\nu}})\right|^2
     \nonumber\\
 && \hspace{5cm}+\tilde{m}_e \tilde{m}_\nu {\rm Re}
      \left\{
       \int_0^\infty
       d^2\widetilde{k}_{e^+} 
       K^2_{i\tilde{\omega}_{e^+}+{1\over2}}(\tilde{\ell }_{e^+})
       \int_0^\infty
       d^2\widetilde{k}_{\bar{\nu}} 
       K^2_{i(\tilde{\omega}_{e^+}-\widetilde{\Delta m})
        -{1\over2}}(\tilde{\ell }_{\bar{\nu}})
      \right\}\Biggr],
\end{eqnarray}
\begin{eqnarray}
 && \Gamma^{p\rightarrow n}_{(\rm iii)} 
   = {2G_F^2\over \pi^7e^{\pi\widetilde{\Delta m}}}
       \int_0^{\widetilde{\Delta m}}
       d\tilde{\omega}_{e^-}
       \Biggl[
       \int_0^\infty
       d^2\widetilde{k}_{e^-} 
       \tilde{\ell }_{e^-} 
 \left|K_{i\tilde{\omega}_{e^-}+{1\over2}}(\tilde{\ell }_{e^-})\right|^2 
       \int_0^\infty
       d^2\widetilde{k}_{\bar{\nu}}  
       \tilde{\ell }_{\bar{\nu}}
 \left|K_{i(\tilde{\omega}_{e^-}-\widetilde{\Delta m})
        +{1\over2}}(\tilde{\ell }_{\bar{\nu}})\right|^2   
     \nonumber\\
 && \hspace{5cm}+\tilde{m}_e \tilde{m}_\nu {\rm Re}
      \left\{
       \int_0^\infty
       d^2\widetilde{k}_{e^-}
       K^2_{i\tilde{\omega}_{e^-}+{1\over2}}(\tilde{\ell }_{e^-})
       \int_0^\infty
       d^2\widetilde{k}_{\bar{\nu}}  
       K^2_{i(\tilde{\omega}_{e^-}-\widetilde{\Delta m})
        -{1\over2}}(\tilde{\ell }_{\bar{\nu}})
      \right\}\Biggr],
\end{eqnarray}
\end{subequations}
By summing them 
we can lump together the integration, and
the total cross section becomes simply in the form
\begin{eqnarray}
 && \Gamma^{p\rightarrow n}_{\rm acc}
  =  \Gamma^{p\rightarrow n}_{(\rm i)}
     +\Gamma^{p\rightarrow n}_{(\rm ii)}
     +\Gamma^{p\rightarrow n}_{(\rm iii)}\nonumber\\
 &&\hspace{9mm}= {2G_F^2\over \pi^7e^{\pi\widetilde{\Delta m}}}
       \int_{-\infty}^\infty 
       d\tilde{\omega}
       \Biggl[
       \int_0^\infty
       d^2\widetilde{k}_e 
       \tilde{\ell }_e 
 \left|K_{i\tilde{\omega}+{1\over2}}(\tilde{\ell }_e)\right|^2 
       \int_0^\infty
       d^2\widetilde{k}_\nu  
       \tilde{\ell }_\nu 
 \left|K_{i(\tilde{\omega}-\widetilde{\Delta m})+{1\over2}}
        (\tilde{\ell }_\nu)\right|^2   
     \nonumber\\
 && \hspace{7cm}+\tilde{m}_e \tilde{m}_\nu {\rm Re}
      \left\{
       \int_0^\infty
       d^2\widetilde{k}_e 
       K^2_{i\tilde{\omega}+{1\over2}}(\tilde{\ell }_e)
       \int_0^\infty
       d^2\widetilde{k}_\nu  
       K^2_{i(\tilde{\omega}-\widetilde{\Delta m})-{1\over2}}(\tilde{\ell }_\nu)
      \right\}\Biggr].\;\; 
\end{eqnarray}
It is hard to deal with these integral of modified Bessel's.
Of course, we can use Eq.~(\ref{k}) again like the case of Section I. 
But we use more useful formula
(see, for example, p219 of Ref.~\cite{HTF} or Ref.~\cite{Gradshteyn}).
\begin{equation}
x^\sigma K_\nu(x)K_\mu(x)= {\sqrt{\pi}\over2}
G^{40}_{24}\left( x^2 \left|
\begin{array}{c}
      {1\over2}\sigma,{1\over2}\sigma+{1\over2} \\\
        {1\over2}(\nu+\mu+\sigma),
        {1\over2}(\nu-\mu+\sigma),
        {1\over2}(-\nu+\mu+\sigma),
        {1\over2}(-\nu-\mu+\sigma)
\end{array}
\right.\right),
\label{kk}
\end{equation}
and by using the definition of $G$ function, 
the integrand can be represented by power of $\ell$ 
and we can easily integrate with respect to $\ell$. 

We find
\begin{eqnarray}
 && \Gamma^{p\rightarrow n}_{\rm acc} 
  ={G_F^2\over 2^3\pi^4e^{\pi\widetilde{\Delta m}}}
 \int_{C_s} {ds\over2\pi i}\int_{C_t} {dt\over2\pi i}
 \int_{-\infty}^\infty d\tilde{\omega}
  {\tilde{m}_e^{2t+1} 
  \tilde{m}_\nu^{2s+1}
  \over
     (2s+1)(2t+1)
     \Gamma(-s+1)
     \Gamma(-t+1) 
  }
  \nonumber\\
&& \hspace{2cm}\times\Biggl[
    \Gamma(-s+{1\over2})
    \Gamma(-t+{1\over2})
    \left|\Gamma(-s+i(\tilde{\omega}-\widetilde{\Delta m})+1)
     \Gamma(-t+i\tilde{\omega}+1)\right|^2     
    \nonumber\\
&& +\tilde{m}_e \tilde{m}_\nu
    \Gamma(-s-{1\over2})
    \Gamma(-t-{1\over2})
     {\rm Re}\left\{
     \Gamma(-s+i(\tilde{\omega}-\widetilde{\Delta m})+1)
     \Gamma(-s-i(\tilde{\omega}-\widetilde{\Delta m}))
     \Gamma(-t+i\tilde{\omega}+1)
     \Gamma(-t-i\tilde{\omega}) 
     \right\}
     \Biggr],
\end{eqnarray}
where all poles of complex $s$ and $t$ planes are picked up 
with $C_s$ and $C_t$, respectively,
by definition of $G$ function.

To integrate with respect to $\omega$, we use the formula of Barnes
 \cite{Gradshteyn}:
\begin{eqnarray}
&& \int_{-i\infty}^{i\infty}d\tilde{\omega}\;
     \Gamma(a+\tilde{\omega})\Gamma(b+\tilde{\omega})
     \Gamma(c-\tilde{\omega})\Gamma(d-\tilde{\omega})
     =2\pi i {\Gamma(a+c)\Gamma(a+d)\Gamma(b+c)\Gamma(b+d)
       \over\Gamma(a+b+c+d)}\nonumber\\[5mm]
&& \hspace{9cm}
   [{\rm Re}\; a,{\rm Re}\; b,{\rm Re}\; c,{\rm Re}\; d>0].
\end{eqnarray}

Eventually, we find that the total cross section in the Rindler frame is 
\begin{eqnarray}
 && \Gamma^{p\rightarrow n}_{\rm acc} 
 ={a^5G_F^2 \over 2^5\pi^{7\over2}e^{\pi\widetilde{\Delta m}}}
    \int_{C_s} {ds\over2\pi i}\int_{C_t} {dt\over2\pi i}
    {
     (\tilde{m}_e^2)^s
     (\tilde{m}_\nu^2)^t
     \over
     \Gamma(-s-t+3)
     \Gamma(-s-t+{7\over2})
    }\nonumber\\
 && \hspace{1.5cm}\times
    \Biggl[
     \left|\Gamma(-s-t+i\widetilde{\Delta m}+3)\right|^2
     \Gamma(-s)
     \Gamma(-t)
     \Gamma(-s+2)
     \Gamma(-t+2)   
     \nonumber\\
 && \hspace{1.5cm}+{\rm Re}\left\{
          \Gamma(-s-t+i\widetilde{\Delta m}+2)
          \Gamma(-s-t-i\widetilde{\Delta m}+4)
        \right\}
     \Gamma(-s+{1\over2})
     \Gamma(-t+{1\over2})
     \Gamma(-s+{3\over2})
     \Gamma(-t+{3\over2})   
     \Biggr].
     \label{eq:10}
\end{eqnarray}
Comparing this to the results in inertial frame~(\ref{result1}), we find that
resulting expression agrees perfectly. This result shows the existence of Unruh
effect is inevitable.

\end{widetext}

\section{Discussions}

We have analyzed the $\beta$ decay and the inverse $\beta$ decay 
of the accelerated proton in both frames. We found analytic expressions for both
frame and found that they agree with each other. If you see the calculation of G.
E. A. Matsas and D. A. T. Vanzella in two-dimensional
model \cite{matsas99,matsas01}, you can realize that on four-dimensional model 
it became hopelessly complicated integral.  
So the main problem in this time is the complication of integral.
To solve it, we used Barnes type representation
as you can see from Eq.~(\ref{k}) and Eq.~(\ref{kk}).
We can demonstrate these formulae by picking up the poles in 
complex plane of the integral valuable.
The sum of these readily becomes the infinity series which
defines the special function.
By using them, we accomplished perfectly analytical proof.

It is straightforward to apply our technique for two dimensional
setup used in Ref.~\cite{matsas99,matsas01},
we can easily prove that the decay rates is independent of the frame.

\appendix

\section{}

In this appendix A we proof both key equations Eq.~(\ref{k}) and Eq.~(\ref{k2}).

Firstly we derive Eq.~(\ref{k}). 
Through the path $C_1$, all poles of $\Gamma(-s)$ and $\Gamma(-s-\mu)$
are picked up (see FIG.1).

\begin{figure}
\includegraphics[width=\linewidth]{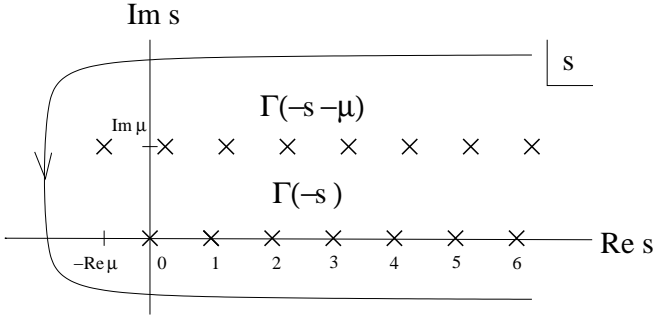}
\caption{\label{fig:epsart}
All residues of $\Gamma(-s)$ and $\Gamma(-s-\mu)$ are picked up.
}
\end{figure}

\begin{eqnarray}
 && {1\over2}\int_{C_1} {ds\over 2\pi i}\Gamma(-s)\Gamma(-s-\mu)
       \left({z\over2}\right)^{2s+\mu}\nonumber\\
 && ={1\over 2} \sum_{n=0}^\infty
     {(-1)^n\over n!}
            \Gamma(\mp\mu -n)\left(z\over 2\right)^{2n\pm\mu}\nonumber\\
 && ={1\over 2} \sum_{n=0}^\infty\sum_\pm
     {(-1)^n\over n!}
     {\pi \over \Gamma(n\pm\mu+1)\sin(-n\mp\mu)\pi}
            \left(z\over 2\right)^{2n\pm\mu}
  \nonumber\\
 && ={\pi\over 2}{-I_\mu(z)+I_{-\mu}(z)\over \sin\mu\pi}\nonumber\\
\end{eqnarray}
where $I_\mu(z)$ is modified Bessel function of the first kind.
The last form is the definition of modified Bessel function 
$K_\mu(z)$ for non-integer $\mu$.
And you find the formula in case $\mu$ is integer $n$ 
by setting $\mu=n$ after partial differentiation of this formula  by $\mu$.

Next we integrate Eq.~(\ref{k2}) as FIG.2.

\begin{figure}
\includegraphics[width=\linewidth]{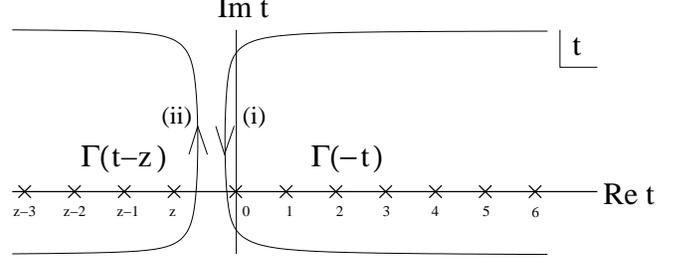}
\caption{\label{fig:epsart}
The contour separates the poles of $\Gamma(-t)$ from those of $\Gamma(t-z)$.}
\end{figure}

\begin{equation}
 \int_{C_2} {dt\over2\pi i}{\Gamma(-t)\Gamma(t-z)\over \Gamma(-z)}
           A^{-t+z}B^t\nonumber
\end{equation}
If you select the contour $(\rm i)$ you find
\begin{eqnarray}
 && =A^z\sum_{n=0}^\infty{(-1)^n\Gamma(-z+n)\over n! \Gamma(-z)}
     \left(B\over A\right)^n\nonumber\\
 && =A^z \left(1+{B\over A}\right)^z\nonumber\\
\end{eqnarray}
So this integration is the expansion form of $(A+B)^z$ in $ B<A$.
Similarly if you select the contour $(\rm ii)$ 
then you obtain the expansion form of $(A+B)^z$ in $ B>A$. 
This integral representation of expansion of $(A+B)^z$ includes
both cases of $ B<A$ and $ B>A$ 
by selecting the contour $(\rm i)$ and $(\rm ii)$,
respectively.

\section{}

In this appendix B, we show the explicit form of the integral.
There are  poles of the power of 1, 2 and 3.

The integral can be simply calculated by the change of variables to
\begin{equation}
  s\rightarrow s-t,\;\; t\rightarrow t.
\end{equation}
 After this transformation we obtain
\begin{widetext}
\begin{eqnarray}
 && \Gamma^{p\rightarrow n} 
 ={a^5G_F^2 \over 2^5\pi^{7\over2}e^{\pi\widetilde{\Delta m}}}
    \int_{C_t}{dt\over2\pi i}
    \int_{C_s}{ds\over2\pi i}
    {
     (\tilde{m}_e^2)^{s-t} 
     (\tilde{m}_\nu^2)^t
     \over
     \Gamma(-s+3)
     \Gamma(-s+{7\over2})
    }\nonumber\\
 && \hspace{1.5cm}\times
    \Biggl[
     \left|\Gamma(-s+i\widetilde{\Delta m}+3)\right|^2
     \Gamma(-t)
     \Gamma(-t+2)
     \Gamma(-s+t)
     \Gamma(-s+t+2)   
     \nonumber\\
 && \hspace{2cm}+{\rm Re}\left\{
          \Gamma(-s+i\widetilde{\Delta m}+2)
          \Gamma(-s-i\widetilde{\Delta m}+4)
        \right\}
     \Gamma(-t+{1\over2})
     \Gamma(-t+{3\over2})
     \Gamma(-s+t+{1\over2})
     \Gamma(-s+t+{3\over2})   
     \Biggr],
     \nonumber\\
\end{eqnarray}
where $\psi(z) = {d\over dz}\ln \Gamma(z)$
and $C_t$ is the path separating the poles of, for example in first term,
$\Gamma(-t)\Gamma(-t+2)$ from those of $\Gamma(-s+t)\Gamma(-s+t+2)$,
and $C_s$ is the path which picks up all the poles in $s$ complex plane.

Firstly, we start by $t$ integration because
the existence of poles in $t$ complex plane is independent of $s$.  
If we integrate by $t$, we obtain for $\tilde{m}_e>\tilde{m}_\nu$:
\begin{eqnarray}
 &&\Gamma^{p\rightarrow n} 
   ={a^5G_F^2 \over 2^5\pi^{7\over2}e^{\pi\widetilde{\Delta m}}}
    \int{ds\over2\pi i}
    {
     \tilde{m}_e^{2s} 
     \over
     \Gamma(-s+3)
     \Gamma(-s+{7\over2})
    } \Biggl\{\Biggl(  
    \Gamma(-s)\Gamma(-s+2)
    +\left({\tilde{m}_\nu\over\tilde{m}_e}\right)^2
    \Gamma(-s+1)\Gamma(-s+3)
    \nonumber\\
 && \hspace{2cm} 
    +\sum_{n=2}^\infty
    \left({\tilde{m}_\nu\over\tilde{m}_e}\right)^{2n}
     \left[\psi(n-1)+\psi(n+1)-\psi(-s+n)-\psi(-s+n+2)
            -2\ln{\tilde{m}_\nu\over \tilde{m}_e}
     \right]
     \nonumber\\
&&  \hspace{7cm}\times
    {\Gamma(-s+n)
     \Gamma(-s+n+2)
\over\Gamma(n-1)\Gamma(n+1)}\Biggr)
      \left|\Gamma(-s+i\widetilde{\Delta m}+3)\right|^2
     \nonumber\\
 && \hspace{2cm}-\sum_{n=0}^\infty
     \left({\tilde{m}_\nu\over\tilde{m}_e}\right)^{2n+3}
     \left[\psi(n+1)+\psi(n+2)-\psi(-s+n+2)-\psi(-s+n+3)
            -2\ln{\tilde{m}_\nu\over \tilde{m}_e}
     \right]
     \nonumber\\
&&  \hspace{5cm}\times{\rm Re}
     \left\{
     \Gamma(-s+i\widetilde{\Delta m}+2)\Gamma(-s-i\widetilde{\Delta m}+4)
     \right\}
     {\Gamma(-s+n+2)
     \Gamma(-s+n+3)
     \over
     \Gamma(n+1)\Gamma(n+2)}
     \Biggr\}
     \nonumber\\ [5mm]
\label{A}
\end{eqnarray}

The simplest case is for massless neutrino.
If we set $m_\nu= 0$, 
then only the first term remains non-vanishing.
After a valuable transformation
$s\rightarrow s+{3\over2}$ we have 

\begin{eqnarray}
 \Gamma^{p\rightarrow n}_{m_\nu=0}
  &=&{a^5G_F^2 \tilde{m}_e^3 \over 2^5\pi^{7\over2}e^{\pi\widetilde{\Delta m}}}
    \int{ds\over2\pi i}
     {\left|\Gamma(-s+i\widetilde{\Delta m}+{3\over2})\right|^2
    \Gamma(-s-{3\over2})
     \Gamma(-s+{1\over2})
\over\Gamma(-s+{3\over2})\Gamma(-s+2)}\tilde{m}_e^{2s} 
     \nonumber\\
 &=& {G_F^2 m_e^3 a^2\over 32\pi^{7\over2}e^{\pi\widetilde{\Delta m}}}
G^{40}_{24}\left( {m_\nu^2\over a^2} \left|
\begin{array}{c}
      {3\over2},\;\;\;2 \\\
        {1\over2},\;\;
       -{3\over2},\;\;
        {3\over2}+i{\Delta m\over a},\;\;
        {3\over2}-i{\Delta m\over a}
\end{array}
\right.\right).
\end{eqnarray}

This is exactly the cross section obtained by Vanzella and Matsas 
with $c_V=1$ and $c_A=0$ (see Eq.~(4.19) in \cite{matsas01_4}).

We perform the integration of Eq.~(\ref{A}) with respect to $s$
and find the following expansions of the cross section

 \begin{eqnarray} 
&& \Gamma^{p\rightarrow n} ={a^5G_F^2 \over \pi^4e^{\pi\widetilde{\Delta m}}}
   \nonumber\\
&& \times \Biggl\{
    \sum_{m=0}^\infty
   \sum_\pm
  \Biggl[\sum_{n=0}^1(-1)^{n+1}
   +\sum_{n=2}^\infty
   {\psi(n-1)+\psi(n+1)-\psi(n-m\mp i\widetilde{\Delta m}-1)
                       -\psi(n-m\mp i\widetilde{\Delta m}-3)
   -2\ln{\tilde{m}_\nu\over\tilde{m}_e}
    \over
  \Gamma(n-1)
  \Gamma(n+1)
    }
  \Biggr]
  \nonumber\\
&& \hspace{2cm}\times
 { (-1)^m
  \Gamma(n-m\mp i\widetilde{\Delta m}-1)
  \Gamma(n-m\mp i\widetilde{\Delta m}-3)
  \over
  (-m\mp2i\widetilde{\Delta m})
  B(m+1,-2m\mp 2i\widetilde{\Delta m})
  }
  \left({\tilde{m}_\nu\over\tilde{m}_e}\right)^{2n}
  \left({\tilde{m}_e\over2}\right)^{2(m\pm i\widetilde{\Delta m}+3)}
  \nonumber\\
&&
  -(1-\gamma)
  \left|\Gamma(i\widetilde{\Delta m}+1)\right|^2
  \left[\left({\tilde{m}_\nu\over 2}\right)^4
       +\left({\tilde{m}_e  \over 2}\right)^4\right]
  +\sum_{k=3}^\infty
  {
  \Gamma(2k-5)
  \left|\Gamma(-k+i\widetilde{\Delta m}+3)\right|^2
  \over
  \Gamma(k-1)
  \Gamma(k+1)
  }
  \left[\left({\tilde{m}_\nu\over 2}\right)^{2k}
       +\left({\tilde{m}_e  \over 2}\right)^{2k}\right]
  \nonumber\\
&&
  -\sum_{k=2}^\infty
  {
  \Gamma(2k-3)
  \left|\Gamma(-k+i\widetilde{\Delta m}+2)\right|^2
  \over
  \Gamma(k-1)
  \Gamma(k+1)
  }
  \left[\left({\tilde{m}_\nu\over 2}\right)^{2k}
        \left({\tilde{m}_e  \over 2}\right)^2
       +\left({\tilde{m}_\nu\over 2}\right)^2
        \left({\tilde{m}_e  \over 2}\right)^{2k}\right]
  \nonumber\\
&& 
   +\sum_{n=2}^\infty
    \sum_{m=2}^\infty
    \Biggl[
   -8\psi(2[n+m-3]+1)
   +2\sum_\pm\psi(-n-m\pm i\widetilde{\Delta m}+3)
   +\sum_\pm\psi(n\pm1)+\sum_\pm\psi(m\pm1)
   -2\ln {\tilde{m}_\nu\tilde{m}_e\over4}\Biggr]\nonumber\\
&&  \hspace{2cm}
    \times
   {
   \Gamma(2[n+m-3]+1)
   \left|\Gamma(-n-m+i\widetilde{\Delta m}+3)\right|^2
    \over
  \Gamma(n-1)
  \Gamma(n+1)
    \Gamma(m-1)\Gamma(m+1)
   }
    \left({\tilde{m}_\nu\over 2}\right)^{2n}
    \left({\tilde{m}_e  \over 2}\right)^{2m}
    \nonumber\\
&& +
    \sum_{n=0}^\infty
    \sum_{m=0}^\infty
   \Biggl(
   -\sum_\pm
  \Biggl[
   \psi(n+1)+\psi(n+2)-\psi(n-m\mp i\widetilde{\Delta m})
                       -\psi(n-m\mp i\widetilde{\Delta m}+1)
   -2\ln{\tilde{m}_\nu\over\tilde{m}_e}
  \Biggr]
  \nonumber\\
&& \hspace{2cm}\times
 { (-1)^m
  \Gamma(n-m\mp i\widetilde{\Delta m})
  \Gamma(n-m\mp i\widetilde{\Delta m}+1)
  \over
  2^2
  \Gamma(n+1)
  \Gamma(n+2)
  (-m\mp 2i\widetilde{\Delta m}+2)
  B(m+1,-2m\mp 2i\widetilde{\Delta m}+2)
  }
  \left({\tilde{m}_\nu\over\tilde{m}_e}\right)^{2n+3}
  \left({\tilde{m}_e\over2}\right)^{2(m\pm i\widetilde{\Delta m}+1)}
  \nonumber\\
 && -
   \sum_\pm
  \Biggl[
   \psi(n+1)+\psi(n+2)-\psi(n-m\mp i\widetilde{\Delta m}-2)
                       -\psi(n-m\mp i\widetilde{\Delta m}-1)
   -2\ln{\tilde{m}_\nu\over\tilde{m}_e}
  \Biggr]
  \nonumber\\
&& \hspace{2cm}\times
 { (-1)^m
  \Gamma(n-m\pm i\widetilde{\Delta m}-2)
  \Gamma(n-m\pm i\widetilde{\Delta m}-1)
  \over
  2^4  
  \Gamma(n+1)
  \Gamma(n+2)
  (-m\pm 2i\widetilde{\Delta m}-2)
  B(m+1,-2m\pm 2i\widetilde{\Delta m}-2)
  }
  \left({\tilde{m}_\nu\over\tilde{m}_e}\right)^{2n+3}
  \left({\tilde{m}_e\over2}\right)^{2(m\mp i\widetilde{\Delta m}+4)}
  \nonumber\\
&& 
  -\Biggl[
    -8\psi(2[n+m]+1)
   +2\sum_\pm\psi(-n-m\pm i\widetilde{\Delta m}\mp1)
   +\psi(n+1)+\psi(n+2)
   +\psi(m+1)+\psi(m+2)-2\ln {\tilde{m}_\nu\tilde{m}_e\over4}\Biggr]\nonumber\\
&&  \hspace{3cm}
    \times
   {
    \Gamma(2[n+m]+1)
    \Gamma(-n-m+i\widetilde{\Delta m}-1)
    \Gamma(-n-m-i\widetilde{\Delta m}+1)
    \over
    \Gamma(n+1)
    \Gamma(n+2)
    \Gamma(m+1)\Gamma(m+2)
   }
    \left({\tilde{m}_\nu\over 2}\right)^{2n+3}
    \left({\tilde{m}_e  \over 2}\right)^{2m+3}
   \Biggr)
   \Biggr\}
   ,\nonumber\\
   \label{a1}
\end{eqnarray}
\end{widetext}
where $B(p,q)={\Gamma(p)\Gamma(q)\over\Gamma(p+q)}$ 
and $\gamma$ is Euler's constant.

This is the final form of the cross section. A natural question is that we have
obtained the result which is not manifestly symmetric with respect to
$\tilde{m}_\nu$ and $\tilde{m}_e$ although the original expression (\ref{result1})
is manifestly symmetric. The resolution of this puzzle is that the integral is of
discontinuous type. Namely, 
 we can obtain the result just by interchanging
$\tilde{m}_\nu$ and
$\tilde{m}_e$ for $\tilde{m}_e<\tilde{m}_\nu$. This can be checked directly by
changing the order of integrations.

\bibliography{decay}

\begin{thebibliography}{15}
\expandafter\ifx\csname natexlab\endcsname\relax\def\natexlab#1{#1}\fi
\expandafter\ifx\csname bibnamefont\endcsname\relax
  \def\bibnamefont#1{#1}\fi
\expandafter\ifx\csname bibfnamefont\endcsname\relax
  \def\bibfnamefont#1{#1}\fi
\expandafter\ifx\csname citenamefont\endcsname\relax
  \def\citenamefont#1{#1}\fi
\expandafter\ifx\csname url\endcsname\relax
  \def\url#1{\texttt{#1}}\fi
\expandafter\ifx\csname urlprefix\endcsname\relax\def\urlprefix{URL }\fi
\providecommand{\bibinfo}[2]{#2}
\providecommand{\eprint}[2][]{\url{#2}}

\bibitem[{\citenamefont{Hawking}(1974)}]{hawking}
\bibinfo{author}{\bibfnamefont{S.~W.} \bibnamefont{Hawking}},
  \bibinfo{journal}{Nature (London)} \textbf{\bibinfo{volume}{248}},
  \bibinfo{pages}{30} (\bibinfo{year}{1974}).

\bibitem[{\citenamefont{Unruh}(1976)}]{unruh76}
\bibinfo{author}{\bibfnamefont{W.~G.} \bibnamefont{Unruh}},
  \bibinfo{journal}{Phys.\ Rev. D} \textbf{\bibinfo{volume}{14}},
  \bibinfo{pages}{870} (\bibinfo{year}{1976}).

\bibitem[{\citenamefont{Fulling}(1973)}]{fulling73}
\bibinfo{author}{\bibfnamefont{S.~A.} \bibnamefont{Fulling}},
  \bibinfo{journal}{Phys.\ Rev. D} \textbf{\bibinfo{volume}{7}},
  \bibinfo{pages}{2850} (\bibinfo{year}{1973}).

\bibitem[{\citenamefont{Davies}(1975)}]{Davies75}
\bibinfo{author}{\bibfnamefont{P.~C.~W.} \bibnamefont{Davies}},
  \bibinfo{journal}{J. Phys. A:\ Gen. Phys.} \textbf{\bibinfo{volume}{8}},
  \bibinfo{pages}{609} (\bibinfo{year}{1975}).

\bibitem[{\citenamefont{Sewell}(1982)}]{sewell82}
\bibinfo{author}{\bibfnamefont{G.~L.} \bibnamefont{Sewell}},
  \bibinfo{journal}{Ann. \ Phys.} \textbf{\bibinfo{volume}{141}},
  \bibinfo{pages}{201} (\bibinfo{year}{1982}).

\bibitem[{\citenamefont{Birrell and Davies}(1982)}]{birrell}
\bibinfo{author}{\bibfnamefont{N.~D.} \bibnamefont{Birrell}} \bibnamefont{and}
  \bibinfo{author}{\bibfnamefont{P.~C.~W.} \bibnamefont{Davies}},
  \emph{\bibinfo{title}{Quantum Field Theory in Curved Space}}
  (\bibinfo{publisher}{Cambridge University Press, Cambridge, England},
  \bibinfo{year}{1982}).

\bibitem[{\citenamefont{Chen and Tajima}(1999)}]{chen99}
\bibinfo{author}{\bibfnamefont{P.}~\bibnamefont{Chen}} \bibnamefont{and}
  \bibinfo{author}{\bibfnamefont{T.}~\bibnamefont{Tajima}},
  \bibinfo{journal}{Phys. Rev.\ Lett.} \textbf{\bibinfo{volume}{83}},
  \bibinfo{pages}{256} (\bibinfo{year}{1999}).

\bibitem[{\citenamefont{Muller}(1997)}]{muller97}
\bibinfo{author}{\bibfnamefont{R.}~\bibnamefont{Muller}},
  \bibinfo{journal}{Phys. \ Rev. D} \textbf{\bibinfo{volume}{56}},
  \bibinfo{pages}{953} (\bibinfo{year}{1997}).

\bibitem[{\citenamefont{et~al.}(1998)}]{caso98}
\bibinfo{author}{\bibfnamefont{C.~Caso} \bibnamefont{et~al.}},
  \bibinfo{journal}{The European Phys. \ Journ. C}
  \textbf{\bibinfo{volume}{3}}, \bibinfo{pages}{1} (\bibinfo{year}{1998}).

\bibitem[{\citenamefont{Matsas and Vanzella}(1999)}]{matsas99}
\bibinfo{author}{\bibfnamefont{G.~E.~A.} \bibnamefont{Matsas}}
  \bibnamefont{and} \bibinfo{author}{\bibfnamefont{D.~A.~T.}
  \bibnamefont{Vanzella}}, \bibinfo{journal}{Phys. \ Rev. D}
  \textbf{\bibinfo{volume}{59}}, \bibinfo{pages}{094004}
  (\bibinfo{year}{1999}).

\bibitem[{\citenamefont{Vanzella and Matsas}(2001{\natexlab{a}})}]{matsas01}
\bibinfo{author}{\bibfnamefont{D.~A.~T.} \bibnamefont{Vanzella}}
  \bibnamefont{and} \bibinfo{author}{\bibfnamefont{G.~E.~A.}
  \bibnamefont{Matsas}}, \bibinfo{journal}{Phys. \ Rev. Lett.}
  \textbf{\bibinfo{volume}{87}}, \bibinfo{pages}{151301}
  (\bibinfo{year}{2001}{\natexlab{a}}).

\bibitem[{\citenamefont{Itzykson and Zuber}(1980)}]{itzykson}
\bibinfo{author}{\bibfnamefont{C.}~\bibnamefont{Itzykson}} \bibnamefont{and}
  \bibinfo{author}{\bibfnamefont{J.-B.} \bibnamefont{Zuber}},
  \emph{\bibinfo{title}{Quantum Field Theory}}
  (\bibinfo{publisher}{McGraw-Hill, New York}, \bibinfo{year}{1980}).

\bibitem[{\citenamefont{Bateman}(1953)}]{HTF}
\bibinfo{author}{\bibfnamefont{H.}~\bibnamefont{Bateman}},
  \emph{\bibinfo{title}{Higher Transcental Functions}}
  (\bibinfo{publisher}{McGRAW-HILL, New York}, \bibinfo{year}{1953}).

\bibitem[{\citenamefont{Gradshteyn and Ryzhik}(1980)}]{Gradshteyn}
\bibinfo{author}{\bibfnamefont{I.~S.} \bibnamefont{Gradshteyn}}
  \bibnamefont{and} \bibinfo{author}{\bibfnamefont{I.~M.}
  \bibnamefont{Ryzhik}}, \emph{\bibinfo{title}{Teble of Integrals, Series and
  Products}} (\bibinfo{publisher}{Academic, New York}, \bibinfo{year}{1980}).

\bibitem[{\citenamefont{Vanzella and Matsas}(2001{\natexlab{b}})}]{matsas01_4}
\bibinfo{author}{\bibfnamefont{D.~A.~T.} \bibnamefont{Vanzella}}
  \bibnamefont{and} \bibinfo{author}{\bibfnamefont{G.~E.~A.}
  \bibnamefont{Matsas}}, \bibinfo{journal}{Phys. \ Rev. D}
  \textbf{\bibinfo{volume}{63}}, \bibinfo{pages}{014010}
  (\bibinfo{year}{2001}{\natexlab{b}}).

\end{thebibliography}

\end{document}